# Empirical calibration for helium abundance determinations in Active Galactic Nuclei

O. L. Dors.[1]⋆, G. C. Almeida[1], C. B. Oliveira[1], S. R. Flury[2,3], R. Riffel[4], R. A. Riffel[5], M. V. Cardaci[6,7], G. F. Hägele[6,7], G. S. Ilha[8], A. C. Krabbe[8], K. Z. Arellano-Córdova[9,10], P. C. Santos[1], I. N. Morais[1]

[1] *Universidade do Vale do Paraíba, Av. Shishima Hifumi, 2911, Zip Code 12244-000, São José dos Campos, SP, Brazil*
[2] *Department of Astronomy, University of Massachusetts Amherst, Amherst, MA 01002, United States*
[3] *NASA FINESST Fellow*
[4] *Departamento de Astronomia, Universidade Federal do Rio Grande do Sul, Av. Bento Gonçalves 9500, Porto Alegre, RS, Brazil*
[5] *Departamento de Física, Centro de Ciências Naturais e Exatas, Universidade Federal de Santa Maria, 97105-900, Santa Maria, RS, Brazil*
[6] *Facultad de Ciencias Astronómicas y Geofísicas, Universidad Nacional de La Plata, Paseo del Bosque s/n, 1900 La Plata, Argentina*
[7] *Instituto de Astrofísica de La Plata (CONICET-UNLP), La Plata, Avenida Centenario (Paseo del Bosque) S/N, B1900FWA, Argentina*
[8] *Departamento de Astronomia, Instituto de Astronomia, Geofísica e Ciências Atmosféricas da USP, Cidade Universitária, 05508-900 São Paulo, SP, Brazil*
[9] *Institute for Astronomy, University of Edinburgh, Royal Observatory, Edinburgh, EH9 3HJ, UK*
[10] *Department of Astronomy, The University of Texas at Austin, 2515 Speedway, Stop C1400, Austin, TX 78712, USA*

**ABSTRACT**

For the first time, a calibration between the He I$\lambda$5876/H$\beta$ emission line ratio and the helium abundance $y=12+\log(\text{He/H})$ for Narrow line regions (NLRs) of Seyfert 2 Active Galactic Nuclei (AGN) is proposed. In this context, observational data (taken from the SDSS-DR15 and from the literature) and direct abundance estimates (via the $T_{\rm e}$-method) for a sample of 65 local ($z < 0.2$) Seyfert 2 nuclei are considered. The resulting calibration estimates the $y$ abundance with an average uncertainty of 0.02 dex. Applying our calibration to spectroscopic data containing only strong emission lines, it yields a helium abundance distribution similar to that obtained via the $T_{\rm e}$-method. Some cautions must be considered to apply our calibration for Seyfert 2 nuclei with high values of electron temperature ($\gtrsim$ 20 000 K) or ionization parameter ($\log U > -2.0$).

## 1 INTRODUCTION

The helium abundance determination in the interstellar medium (ISM) is paramount for imposing stringent constraints on models of the early Universe (e.g. Planck Collaboration 2020) and on chemical evolution models of galaxies (e.g. Carigi & Peimbert 2008) as well as playing a significant role in investigating stellar nucleosynthesis (e.g. Woosley 2019).

Over decades, the abundance ratio of helium to hydrogen has mainly been derived in star-forming regions (SFs, i.e. H II regions and H II galaxies; e.g. Izotov et al. 1999; Hägele et al. 2006, 2008, 2012; Valerdi et al. 2021a) and some (relatively) few studies have focused in Active Galactic Nuclei (AGN, e.g. Bahcall & Oke 1971; Baldwin 1975; Dors et al. 2022). The total helium abundance in relation to hydrogen $y = N(\text{He})/N(\text{H})$, with $N$ being the abundance in number of atoms, in the ISM of galaxies is considered to be

$$y = y^0 + y^+ + y^{2+}, \tag{1}$$

where $y^0 = N(\text{He}^0)/N(\text{H}^+ + \text{H}^0)$, $y^+ = N(\text{He}^+)/N(\text{H}^+ + \text{H}^0)$ and $y^{2+} = N(\text{He}^{2+})/N(\text{H}^+ + \text{H}^0)$[1]. The ionic abundance ratio $y^+$ and $y^{2+}$ are derived from the He I$\lambda$5876/H$\beta$ and He II$\lambda$4686/H$\beta$ line intensity ratios, respectively, and assuming specific values for the electron temperature ($T_{\rm e}$) where the He$^+$ and He$^{2+}$ ions are located (e.g. Berg et al. 2021). The determination of the neutral fraction $y^0$ is indirect. It requires the employment of an Ionization Correction Factor (ICF) relying either on photoionization models or on the empirical relation (He$^0$)/(He)=(S$^+$)/(S) (Peimbert et al. 1992). These approaches indicate that SFs and AGN (Seyfert 2) have around 3% (e.g. Méndez-Delgado et al. 2022) and 50% (e.g. Dors et al. 2022), respectively, of helium in the neutral stage.

For precise helium abundance derivation, for instance, the estimation of the primordial $y_{\rm P}$ abundance (Peimbert & Torres-Peimbert 1974), it is mandatory (e.g. Hägele et al. 2008) to apply the $T_{\rm e}$-method[2]. This method requires measurements of temperature-sensitive nebular features, for instance, the [O III]$\lambda$4363 and [N II]$\lambda$5755 auroral emission lines, which are generally weak (about 100 times weaker than H$\beta$) in objects with high metallicity and/or low ionization stage (e.g. Dors et al. 2008). The $T_{\rm e}$-method was originally designed for H II region and Planetary Nebulae spectra (Peimbert & Costero 1969). However, Dors et al. (2020) provided an adaptation of the $T_{\rm e}$-method for Narrow Line Regions (NLRs) of AGN, permitting to estimate abundances of some heavy elements (O, Ne, Ar, S) in a sample of local ($z < 0.4$) objects (e.g. Armah et al. 2021; Monteiro & Dors 2021; Dors et al. 2023). It is worthing to emphasize that some cautions must be considered in the applicability of the $T_{\rm e}$-method for AGN, such as the presence of shocks that produce very high electron temperature values ($T_{\rm e} > 20\,000$ K, e.g. Riffel et al. 2021), precluding the use of this method.

---

⋆ E-mail: olidors@univap.br (OLD)

[1] For simplicity, throughout this paper, the $N$ term is not used.

For the cases where it is not possible to apply the $T_{\rm e}$-method, a plethora of so-called 'strong-line methods', suggested by Pagel et al. (1979), have been employed mainly to derive the O/H abundance, which is traditionally assumed as a metallicity ($Z$) tracer in SFs (e.g. Sanders et al. 2024). Hereafter, Storchi-Bergmann et al. (1998) proposed a first theoretical calibration between strong optical emission line ratios and the O/H abundance of NLRs (see also Castro et al. 2017; Carvalho et al. 2020; Dors et al. 2021; Dors 2021). Some reasons can explain why oxygen is largely assumed as a $Z$ tracer (e.g. Armah et al. 2023), for instance, not only due to the strong emission lines ([O II]$\lambda$3727, [O III]$\lambda$5007) emitted by its most abundant ions (O$^+$ and O$^{2+}$, e.g. see Flury & Moran 2020) present in the optical spectrum of line emitter objects but also, oxygen is the most abundant heavy element in most astrophysical regions and is ejected promptly after star formation.

Calibrations between strong emission lines and abundances for other elements (e.g. N, S) are barely found in the literature. For instance, Díaz & Pérez-Montero (2000) proposed a calibration between the sulfur lines and the S/H abundance. While studies have mainly focused on strong-line methods relying on collisionally excited lines, recombination lines have received little attention. So far, it seems that the unique calibrations for SFs involving recombination lines, as those measured by Esteban et al. (2002, 2005, 2009, 2014), are the ones proposed by Peimbert & Peimbert (2006) and Valerdi et al. (2021b), the latter proposing a calibration between He I$\lambda$5876/H$\alpha$ and $y^+$.

In the present study, we take advantage of direct (more precise) estimates of helium abundances in the NLRs of Seyfert 2 nuclei presented by Dors et al. (2022) to propose, for the first time, an empirical calibration between the He I$\lambda$5786/H$\beta$ line ratio and $y$, for AGN. The He I$\lambda$5786 line is the brightest He I recombination line in the optical wavelengths, being (relatively) easy to measure in SFs and AGN spectra. Indeed, recent JWST/NIRSpec observations have yielded significant He I measurements in galaxies at $z > 5$ (e.g. Bunker et al. 2023; Yanagisawa et al. 2024), which prioritized "rare" galaxies including AGN. While most studies aim to determine the primordial helium abundance and therefore seek a highly accurate determination, here, we proposed a method to estimate $y$ with similar uncertainty to that in O/H calibrations. This Letter is organized as follows. In Section 2, we describe the methodol-ogy used to obtain the calibration. Sect. 3 presents the results and discussions, while the conclusions are given in Sect. 4.

---

[2] For a review of the $T_{\rm e}$-method see Peimbert et al. (2017) and Pérez-Montero (2017).

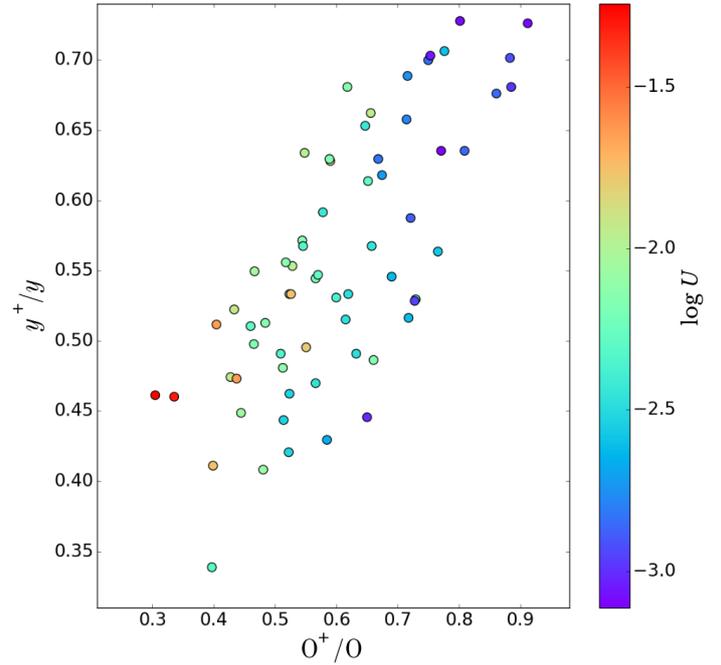

**Figure 1.** Abundance ratio of ($y^+/y$) versus (O$^+$/O) for NLRs of AGN obtained by Dors et al. (2022) using the $T_{\rm e}$-method. The color bar indicates the scale of the ionization parameter ($U$) derived from its relation with [O III]$\lambda$5007/[O II]$\lambda$3727 proposed by Carvalho et al. (2020).

## 2 METHODOLOGY

To obtain a calibration between the He I$\lambda$5876/H$\beta$ line intensity ratio and $y = $ He/H for NLRs of Seyfert 2, we used the observational emission-line intensities and chemical abundance estimates (through the $T_{\rm e}$-method) by Dors et al. (2022). The observational data comprise two sets of narrow optical emission lines [full-width at half-maximum (FWHM) lower than 1000 km s$^{-1}$]. The first sample (10 objects) is taken from SDSS-DR15 database (Aguado et al. 2019), where each spectrum was corrected for Galactic extinction using the Cardelli et al. (1989) law. Additionally, the underlying stellar population – fitted with the STARLIGHT code (Cid Fernandes et al. 2005) – was subtracted, resulting in the pure emission line spectrum. The emission lines were measured using the publicly available IFSCUBE package (Ruschel-Dutra et al. 2021). The second sample (55 objects) comprises emission line intensities, corrected by reddening, obtained by different authors, and taken from the literature. Since several measurements for emission lines compiled from the literature do not present their uncertainties, we adopted them in our study with a typical error of 10% for strong emission lines and an error of 20% for weak emission lines. It is worth mentioning that the sample compiled from the literature is heterogeneous, where emission lines were corrected by extinction by the authors assuming different approaches. However, Dors et al. (2022) showed that the uncertainty produced in the intensity of the [O II]$\lambda$3727/H$\beta$ line ratio (strongly dependent on the reddening) due to the use of distinct approaches to reddening correction is in order of that produced by the error line measurements ($\sim 0.1$ dex).

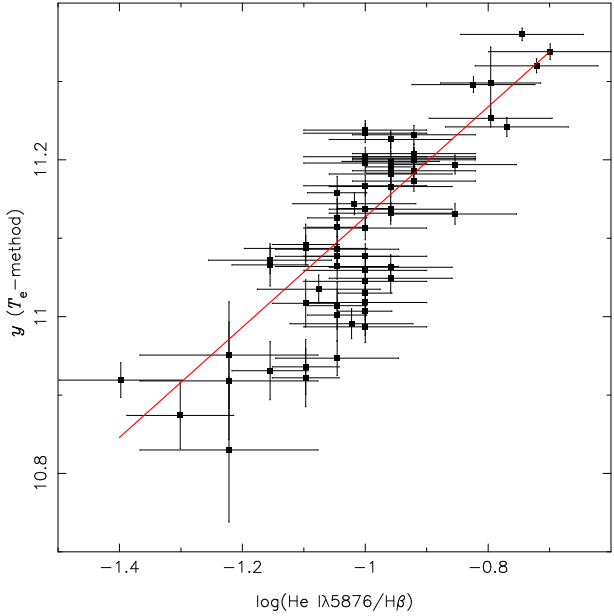

**Figure 2.** Abundance of y=12+log(He/H) versus the logarithm of He I$\lambda$5876/H$\beta$ line ratio. Points represent abundance estimates via the $T_e$-method and observational data of Seyfert 2 nuclei (see Sect. 2), both taken from Dors et al. (2022). Line represents the fitting to the points given by the Eq. 2. Error bars represent the uncertainty in the observational data and abundance estimates.

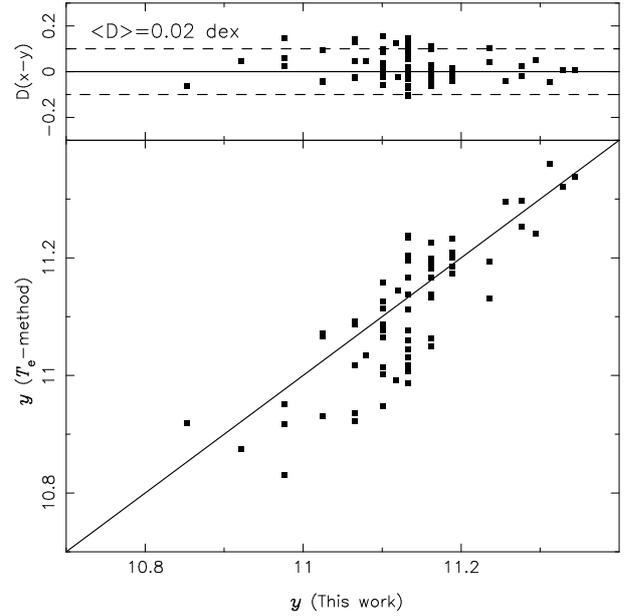

**Figure 3.** Bottom panel: Comparison between helium abundances [in units of 12+log(He/H)] computed using the Eq. 2 and those derived through the $T_e$-method by Dors et al. (2022). Solid line represents the equality between them. Top panel: difference (D = x–y) between both estimations. Line represents the null difference between the estimates, while dashed lines represent the maximum uncertainty ($\pm 0.1$ dex) derived for y via the $T_e$-method. The average difference ($< D >$) is indicated.

After the emission-line reduction/compilation is carried out, only the objects classified as AGN, according to the criteria proposed by Kewley et al. (2001) to separate SF and AGN objects and by Cid Fernandes et al. (2010) to separate AGN-like and Low-ionization nuclear emission-line region (LINER) objects were considered. The final sample results in 65 Seyfert 2 nuclei with redshift $z \lesssim 0.2$. The emission line intensities are reported in Table A1 of Dors et al. (2022).

The helium abundances were calculated by Dors et al. (2022) by using the PyNeb code (Luridiana et al. 2015) and adopting the $T_e$-method. Briefly, the following methodology was assumed: (i) electron temperature for the high ionization zone $T_{\rm high}$ (assumed in the $y^{2+}$ derivation) was calculated from its relation with the [O III]$\lambda(4959 + \lambda 5007)/\lambda 4363$ line ratio; (ii) electron temperature for the low ionization zone $T_{\rm low}$ (assumed in the $y^+$ derivation) was estimated (indirectly) by using the theoretical relation with $T_{\rm high}$ proposed by Dors et al. (2020); (iii) the $y^+$ and $y^{2+}$ ionic abundances were calculated from the He I$\lambda$5876/H$\beta$ and He II$\lambda$4686/H$\beta$ line intensity ratios, respectively, and assuming electron density values derived, for each object, from the observational [S II]$\lambda$6716/$\lambda$6731 line ratio; (iv) to estimate $y^0$, a theoretical ICF proposed by Dors et al. (2022) and relied on the results of a grid of photoionizations models computed with the CLOUDY code version 17.00 (Ferland et al. 2013) was adopted; and (v) the total helium abundance was derived through Equation 1.

The final sample spans a wide range of helium abundances $(y/y_\odot) = 0.6 - 2.5$, metallicities $[(Z/Z_\odot) = 0.3 - 2.6]$ and ioniza-tion parameters $(-3.8 \lesssim \log U \lesssim -0.6)$[3]. The observational average error in the He I$\lambda$5876/H$\beta$ line ratio is $\sim 15\%$. The average error in y derivations is in the order of 0.02 dex (see also Hägele et al. 2006, 2008), with this error reaching up to ∼ 0.1 dex for objects with the lowest He ɪ$\lambda$5876/H$\beta$ line ratio intensities (see below).

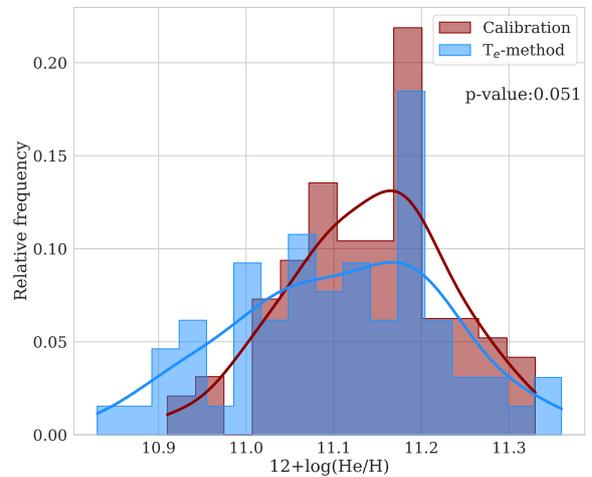

**Figure 4.** Histogram containing the y=12+log(He/H) abundance distributions for Seyfert 2 galaxies. Red distribution is based on values derived from our empirical calibration (Eq. 2) applied to a new sample (data from SDSS-DR15) of 104 Seyfert 2 nuclei. Blue distribution is based on values derived via the $T_e$-method for the sample of 65 Seyfert 2 nuclei described in Sect. 2 and estimated by Dors et al. (2022).

---
[3] Values obtained from the calibration between [O III]$\lambda$5007/[O II]$\lambda$3727 line ratio and $U$ proposed by Carvalho et al. (2020).

## 3 RESULTS & DISCUSSION

In general, abundance calibrations consider emission line intensities emitted by the most abundant ions of a given element, for instance, the calibration between the $R_{23}$=([O ɪɪ]$\lambda$3727+[O ɪɪɪ]($\lambda$4959 + $\lambda$5007))/H$\beta$ line ratio and the O/H abundance (Pagel et al. 1979). In the present study, we proposed a calibration relying on a line emitted by a unique ion, i.e. He$^+$, whose abundance could represent a small fraction of the total helium abundance, resulting in a marginal calibration. In Figure 1, direct estimates of ($y^+/y$) versus (O$^+$/O) (a tracer of the ionization degree) for NLRs from Dors et al. (2022) are shown. In this figure, the color bar indicates the ionization parameter ($U$) scale as derived from its relation with the [O ɪɪɪ]$\lambda$5007/[O ɪɪ]$\lambda$3727 line ratio proposed by Carvalho et al. (2020). We can see that the fraction of He$^+$ in relation to the total helium abundance ranges from ∼ 40% (for objects with high $U$) to ∼ 75% (for objects with low $U$). Thus, for AGNs with high ionization parameters (i.e. $\log U \gtrsim -2.0$) our empirical calibration can produce uncertain helium abundance. For these cases, we suggest either using an accurate ICF estimate or measuring the other ionization states directly if possible. In any case, Carvalho et al. (2020), through a comparison between results of photoionization models and observational data taken from the SDSS DR7 data set, found that most NLRs of AGNs in the local universe present ionization parameters in the range $-4.0 \lesssim \log U \lesssim -2.5$.

In Fig. 2, the resulting calibration is shown, where a strong positive correlation ($R = 0.83$) between $y$ and log(He ɪ$\lambda$5876/H$\beta$) is achieved. To derive an expression for the (He ɪ$\lambda$5876/H$\beta$)-$y$ relation, we performed 1000 bootstrap realizations with the Huber Regressor model (Owen 2007), taking into account the errors in both axes. The resulting fitting to the points is given by

$$12 + \log(\text{He/H}) = 0.703(\pm 0.05)\, x + 11.83(\pm 0.05), \quad (2)$$

being $x$ =log(He ɪ$\lambda$5876/H$\beta$).

To analyze the reliability of our calibration, in Fig. 3 (bottom panel), we compare the helium abundances for the objects in our AGN sample (see Sect. 2) derived through the Eq. 2 with those via the $T_e$-method from Dors et al. (2022), where a good agreement between both estimates can be noted. In Fig. 3 (top panel) the difference (D) between the estimates versus the $y$ abundances via Eq. 2 is shown. We find no systematic behavior between D and $y$, being the difference (< D >) equal to 0.02 dex, lower than the uncertainty attributed (∼ 0.2 dex) to strong-line methods (see e.g. Denicoló et al. 2002).

As an additional test of our calibration, we consider a larger AGN sample than the one in the present study and derived the helium abundance distribution through our calibration, i.e. Eq. 2. The AGN observational data for this larger sample are obtained from the SDSS-DR15 by adopting the criterion proposed by Molina et al. (2021), from which objects with log(He ɪɪ $\lambda$4686/H$\beta$) $\gtrsim -1.0$ are classified as AGN, otherwise, SFs. The same data reduction procedures described in Sect. 2 are applied to this new sample resulting in 104 Seyfert 2 nuclei located in the local universe ($z < 0.4$). We emphasize that the larger sample does not contain a similar population of galaxies to the calibration sample, and the He/H abundances for both samples are used as a simple test of our calibration rather than as evidence for its accuracy. In Fig. 4 the $y$ distributions derived by using our calibration (Eq. 2) for the larger AGN sample (red distribution) and via the $T_e$-method for the Dors et al. (2022) sample (blue distribution) are shown. Despite the discrepancy between the two distributions at low He/H abundances, one can see a good agreement between them, having very similar 12+log(He/H) mean values: 11.14±0.09 dex and 11.11±0.11 dex, for the larger and the Dors et al. (2022) samples, respectively. Also, the p-value from the Anderson-Darling test is 0.051, indicating that both data samples follow a similar distribution.

## 4 CONCLUSION

We present the first calibration between the He ɪ$\lambda$5876/H$\beta$ emission line ratio and the $y$=12+log(He/H) abundance ratio for Narrow Line Regions (NLRs) of Seyfert 2 Active Galactic Nuclei (AGN). We used narrow optical emission line intensities taken from the SDSS-DR15 and from the literature, and direct estimates (based on the $T_e$-method) for a sample of 65 Seyfert 2 nuclei ($z < 0.2$). The resulting calibration yields $y$ abundance with uncertainties in order of 0.02 dex, lower than those (∼ 0.2 dex) attributed to strong-line methods proposed to estimate the O/H abundance. Applying our calibration to a larger observational data sample of 104 objects containing only strong emission lines, we find a similar helium abundance distribution to that obtained via the $T_e$-method. We emphasize that some cautions must be considered to apply our calibration for AGNs with high electron temperature values ($\gtrsim$ 20 000 K), in which a high contribution of shocks to the heating/ionization of the gas is expected. Moreover, somewhat uncertain abundance values for Seyfert 2 nuclei with high ionization parameter ($\log U > -2.0$), whose He$^{2+}$ abundance could represent a large fraction of the total helium abundance, are derived from our (He ɪ/H$\beta$)-$y$ relation.

## ACKNOWLEDGEMENTS

OLD is grateful to Fundação de Amparo à Pesquisa do Estado de São Paulo (FAPESP), process number 2022/07066-6, and to Conselho Nacional de Desenvolvimento Científico e Tecnológico (CNPq). GCA and INM are grateful to Coordenação de Aperfeiçoamento de Pessoal de Nível Superior (CAPES). CBO is grateful to FAPESP for the support under grant 2023/10182-0. GSI acknowledges financial support from FAPESP Proj. 2022/11799-9.

## DATA AVAILABILITY

The data underlying this article will be shared on reasonable request to the corresponding author.

This paper has been typeset from a TeX/LaTeX file prepared by the author.